\documentclass[final,3p,times,onecolumn]{elsarticle}
\usepackage{amsmath,amssymb,amsfonts,scalerel}
\usepackage{blindtext}
\usepackage{graphicx}
\usepackage{adjustbox}
\usepackage{epsfig}
\usepackage{float}
\usepackage{subfig}
\usepackage{xcolor}
\usepackage[colorlinks=true,citecolor=blue,linkcolor=blue,urlcolor=blue]{hyperref}
\usepackage{setspace}
\usepackage{multirow}
\usepackage{multicol}
\usepackage{url}
\usepackage{hyperref}
\usepackage{rotating}
\usepackage{geometry}
\biboptions{sort,compress}
\journal{AIP Advances}

\begin{document}

\begin{frontmatter}

\title{DFT + U Study of structural, electronic, optical and 
       magnetic properties of LiFePO$\rm_{4}$ Cathode 
       materials for Lithium-Ion batteries}

\author{A.K.~Wabeto}
\author[]{K.N.~Nigussa\corref{cor1}}
\cortext[cor1]{Corresponding author:\ kenate.nemera@aau.edu.et\ 
               (K.N.~Nigussa)}
\author[]{L.D.~Deja}
\ead{lemi.demeyu@aau.edu.et}

\address{Department of Physics,\ Addis Ababa University,\ 
         P.O. Box 1176,\ Addis Ababa,\ Ethiopia}
         
\begin{abstract}
In this study,\ we have employed\ a DFT+U calculation using\ 
quantum-espresso (QE)\ code to investigate\ the structural,\ 
electronic,\ optical,\ and magnetic\ properties of\ 
LiFePO$\rm_{4}$\ cathode material\ for Li-ion batteries.\ 
Crystals of LiFePO$\rm_{4}$\ and related\ materials have\ 
recently\ received\ a lot of\ attention due to\ their\ 
very promising\ use as cathodes\ in rechargeable\ 
lithium-ion batteries.\ The structural optimization\ 
was performed\ and the equilibrium\ parameters such\ 
as the lattice\ constants,\ and the bulk modulus\ 
are calculated\ using QE code\ and found to be\ 
a=4.76~$\rm{\AA}$,\ b=6.00~$\rm{\AA}$,\ c=10.28~$\rm{\AA}$,\ 
$\beta$ = 90.2~GPa,\ respectively.\ The projected\ 
density\ of states (PDOS)\ for the LiFePO$\rm_{4}$\ 
material is\ remarkably similar\ to experimental\ 
results\ in\ literature\ showing\ the transition\ 
metal 3$d$\ states\ forming narrow\ bands above\ 
the O 2$p$\ band.\ The results\ of the various spin\ 
configurations suggested\ that the ferromagnetic\ 
configuration can serve\ as a useful approximation\ 
for studying\ the general features\ of these systems.\ 
In the\ absence of Li,\ the majority spin\ transition\ 
metal 3$d$\ states are\ well-hybridized with the\ 
O $2p$ band in FePO$\rm_{4}$.\ The result obtained\ 
with a DFT + U\ showed\ that LiFePO$\rm_{4}$\ 
is direct\ band gap materials\ with a band gap of\ 
3.82~eV,\ which is within\ the range of the experimental\ 
values.\ The PDOS\ analyses show\ qualitative information\ 
about the\ crystal field\ splitting and bond\ hybridization\ 
and help\ rationalize the\ understanding of the\ structural,\ 
electronic,\ optical,\ and magnetic properties of the\ 
LiFePO$\rm_{4}$\ as a novel\ cathode material.\ On the\ 
basis of\ the predicted optical\ absorbance,\ reflection,\ 
refractive index,\ and\ energy loss function,\ LiFePO$\rm_{4}$\ 
is demonstrated\ to be viable and cost-effective,\ 
which is very\ suitable as\ a cathode material\ for\ 
Li-ion battery.\
\end{abstract}


\end{frontmatter}

\section{Introduction\label{sec:intro}}

Energy storage\ is a critical problem\ in the 21$\rm^{st}$\ 
century.\ As the world\ population grows,\ so too does the\ 
demand for\ energy and energy\ storage materials.\ The\ 
development of the\ next generation\ of cars,\ personal\ 
electronics,\ and renewable energy\ sources hinges on\ 
improvements in\ battery technology.\ Batteries are\ 
one of the\ most promising energy\ storage technologies\ 
due to their\ high conversion efficiency\ and essentially\ 
zero emissions~\cite{joseph2006battery}.\ Lithium-ion\ 
batteries (LIBs)\ are considered to be\ one of the most\ 
promising batteries\ owing to their\ high power density,\ 
long cycle life,\ and environmental friendliness,\ which\ 
leads to\ their wide use\ in portable electronic devices~\cite{arevalo2021novel}.\
The cathode material\ is the most\ crucial component of\ 
LIBs.\ Therefore,\ tremendous efforts\ have been dedicated\ 
to the\ development of\ cathode materials.\ The cathode\ 
materials of LIBs\ are usually\ intercalation compounds,\ 
including layered\ LiMO$\rm_{2}$,\ (M=Co, Ni, Mn),\ 
LiNi$\rm_{1-x}Co\rm_{x}Mn\rm_{y}$,\ spinel\ LiMn$\rm_{2}O_{4}$\ 
and olivine\ LiFePO$\rm_{4}$ materials.\ Among them,\ 
olivine-structured\ LFP was proposed in 1997 by\ 
Pandhi~\cite{figgener2021development}\ with excellent\ 
cycling stability,\ low cost,\ and good safety.\ 
Nevertheless,\ the poor ionic\ and electronic conductivity\ 
and low Li$\rm^+$\ diffusion has hindered\ its further\ 
application.\ Morphology control,\ particle size reduction,\ 
surface coatings,\ and cation or\ anion doping have been\ 
applied to\ improve its\ properties~\cite{hassan2017optimal}.\ 
Furthermore,\ it is significant\ to explicitly\ understand\ 
the microscopic\ origins of\ these improvements.\\

\begin{figure}
\centering
\includegraphics[scale=0.5]{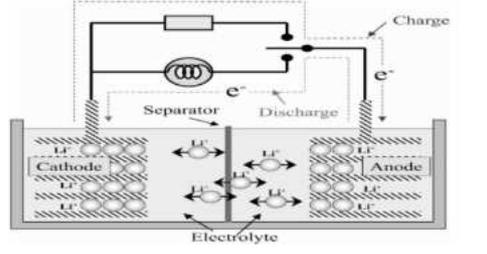} 
\caption{Schematic representation of a common Li-ion battery.
\label{fig1}}
\end{figure}

Lithium iron\ phosphate,\ which is\ an ordered type\ 
compound,\ is under\ extensive studies\ as one of the\ 
most promising\ cathode material.\ It has more favored\ 
properties such\ as low cost,\ environmental compatibility,\ 
less toxicity,\ high thermal stability,\ and high specific\ 
capacity\ compared to the\ LiCo$\rm_{2}$ and LiMn$\rm_{2}O\rm_{4}$.\ 
Most exciting\ advantages of LiFePO$\rm_{4}$\ is its stability\ 
with high\ voltage application,\ hardly changes\ while Li-ion\ 
intercalation\ and deintercalation.\ In Li-ion battery,\ 
lithium ions\ are extracted from\ anode to cathode during\ 
discharge process\ and it is reversed\ during charging as\ 
depicted in Fig~\ref{fig1}.\ The extraction\ and insertion\ 
of lithium during\ charging and\ recharging process\ may be\ 
written as Eqs.~\eqref{eq1}~$\&$~\eqref{eq2}\ below.\ However,\ 
the poor intrinsic\ electronic and ionic\ conductivities\ 
of LiFePO$\rm_{4}$\ limits its\ practical use.\ Moreover,\ 
the band-gap\ of LiFePO$\rm_{4}$\ is under debate\ which\ 
requires more\ structural and\ electronic analysis~\cite{saw2014electro, hassoun2014advanced}.\\

\begin{equation}{\label{eq1}}
LiFePO\rm_{4} - xLi\rm^{+} - xe\rm^{-} 
{\rightarrow} FePO\rm_{4} + (1-x)LiFePO\rm_{4}       
\end{equation} 

\begin{equation}{\label{eq2}}
FePO\rm_{4} - xLi\rm^{+} - xe\rm^{-} 
{\rightarrow} LiFePO\rm_{4} + (1-x)LiFePO\rm_{4}       
\end{equation} 

Even though\ the low cost,\ good stability,\ and competitive\ 
electrochemical properties\ make the olivine\ Li$\rm_{x}MPO\rm_{4}$\ 
family an\ exciting new area\ for cathode development\ 
in Li\ rechargeable batteries,\ they are facing challenges\ 
due to their\ low electrical conductivity.\ Among these\ olivine\ 
cathode materials,\ LiFePO$\rm_{4}$,\ which is in its pure\ 
form has\ very poor conductivity,\ greatly inhibited\ 
high-rate applications.\ Efforts to\ increase conductivity\ 
of electrodes\ made from\ the materials\ have focused\ 
on particle\ size reduction,\ intimate carbon coating,\ 
and cation doping.\ Significant disagreement\ exits on\ 
the origin\ of the low\ electronic\ conductivity~\cite{omar2012rechargeable}.\\
Ab-initio studies\ focusing on the\ band gap and\ 
effective hole\ or electron mass\ have found\ a small gap,\ 
or no gap\ at the Femi level,\ which seems\ to be in\ 
contradiction to\ experiment.\ However,\ there is significant\ 
evidence that\ the local density\ approximation (LDA)\ 
and generalized\ gradient approximation\ (GGA) used in\ 
almost all\ previous studies\ on the electronic\ structure\ 
of these\ phosphates cannot\ accurately reproduce\ their\ 
electronic structure\ due to the\ approximate treatment\ 
of the electron\ correlation in\ transition metal\ 
orbitals by LDA/GGA.\ In order to\ clarify the\ 
electronic structure\ of LiFePO$\rm_{4}$\ we have applied\ 
the more\ accurate DFT+U\ (LDA+U/GGA+U) method\ to determine\ 
the projected\ density of states\ (PDOS) of these\ 
materials~\cite{satyavani2016methods}.\\

Concerns with the\ safety,\ cost,\ charge/discharge rates,\ 
cycle life,\ and energy density\ of Li-ion batteries\ 
represent the\ main challenges\ in Li-ion development.\ 
Additionally,\ if Li-ion batteries\ are to be employed\ 
in Hybrid Electrical\ Vehicles (HEVs)\ then gravimetric\ 
energy density,\ uniformity in\ performance of individual\ 
cells inside\ a complex,\ multicell battery,\ and cost\ 
are the fields\ where more\ research is absolutely\ 
necessary~\cite{xiao2016electrical}.\
Thus,\ current rapid\ development of\ society requires\ 
a major advancement\ in the battery materials\ to achieve\ 
high accuracy,\ long life cycle,\ low cost,\ and reliable\ 
safety.\ Therefore,\ many new efficient\ energy storage\ 
materials\ and battery systems\ are being developed and\ 
explored,\ and their working mechanisms\ must be clearly\ 
understood\ before industrial\ applications~\cite{alfaruqi2020density}.\	
Nowadays,\ computers are\ very useful tools\ for condensed\ 
matter physics\ and material\ sciences and\ they have been\ 
used to\ predict the electronic,\ optical,\ and magnetic\ 
properties\ of materials by\ using a suitable\ computing\ 
method~\cite{oukahou2022investigation}.\ By now,\ a lot of\ 
first principle\ calculations\ have been\ performed on\ 
LiFePO$\rm_{4}$\ cathode materials\ and FePO$\rm_{4}$~\cite{he2019density}\ 
to analyze\ electronic,\ optical,\ and magnetic properties\ 
of LiFePO4.\ The focus of\ this study was\ the electronic\ 
structure calculation\ and analysis of\ optical,\ and magnetic\ 
properties for\ LiFePO$\rm_{4}$\ and end\ material of\ 
LiFePO$\rm_{4}$,\ FePO$\rm_{4}$\ within density\ functional\ 
theory (DFT)\ frame work.\ From many aspects,\ iron is an\ 
attractive metal\ for use in the\ field of battery materials\ 
since it is\ abundant and\ environmentally friendly~\cite{nakayama2016density}.\ 
Crystals of\ LiFePO$\rm_{4}$\ and related materials\ have\ 
recently received\ a lot of attention\ due to their\ 
very promising\ use as cathodes\ in rechargeable\ 
lithium ion\ batteries~\cite{saal2013materials}.\\

The paper is organized as follows.~In\ 
the\ next\ section (sec.~\ref{sec:comp}),\ 
a\ detail account\ of the computational\ 
method\ is presented.\ Results\ and\ 
discussion are\ presented in\ 
section~\ref{sec:res},\ with\ the\ 
conclusion\ being presented in\ 
section~\ref{sec:conc}.
\section{Computational Methods\label{sec:comp}}
An ab-initio simulations within quantum 
espresso\ code~\cite{Giannozzietal2009}\
is\ used\ to examine the\ electronic structure\ 
and optical properties of LiFePO$\rm_{4}$.\ 
The\ electron wave-function\ is expanded over\ a plane\ 
wave basis\ set.\ The\ electron-ion\ interactions\ 
is\ approximated\ within\ projector augmented\ 
wave~(PAW) modality~\cite{PB94}\ 
upon\ the\ calculation\ of\ electronic\ properties\ 
and geometry\ optimization.\ Upon\ optical\ 
properties\ calculations,\ the\ electron-ion\ 
interactions\ is\ approximated\ within\ norm\ 
conserving\ pseudopotential~\cite{KB82}.\  
The exchange-correlation\ energies are\ treated\ 
using\ PBE~\cite{PBE96}.\
The\ k-points\ of\ the\ Brillouin zone~(BZ)\ are\ 
generated from the input ${\bf{k}}$-mesh\ using\ 
the\ Monkhorst-Pack scheme~\cite{MP76}.\

The\ number\ of valence electrons\ considered\ for\ 
each element\ within the\ paw\ data\ sets\ is\ Li:1,\ 
Fe:8,\ P:5,\ and O:6.\ Geometry\ relaxations are\ 
carried out using BFGS\ minimizer~\cite{BS82},\ 
where optimization\ of\ the atomic\ coordinates\ 
and the unit\ cell degrees of freedom\ is done\ 
within\ the concept\ of the Hellmann-Feynman forces\ 
and\ stresses~\cite{PRF39, NM85}\ as\ calculated on\ 
the Born-Oppenheimer~(BO) surface~\cite{WM91}.\ 
The convergence\ criteria for the forces\ were set\ 
at 0.05 eV/{\AA}.\
A\ van\ der\ Waal's\ treatment\ within\ DFT-D3~\cite{GEG2011}\ 
is\ applied\ wherever\ necessary.\ 
The {\bf{k}}-mesh of 4{$\times$}4{$\times$}4\ 
and a\ cut-off\ energy~(ecut)~of\ 600 eV is\ 
used\ in\ the\ calculations.\

\begin{figure}
\centering
\includegraphics[scale=0.5]{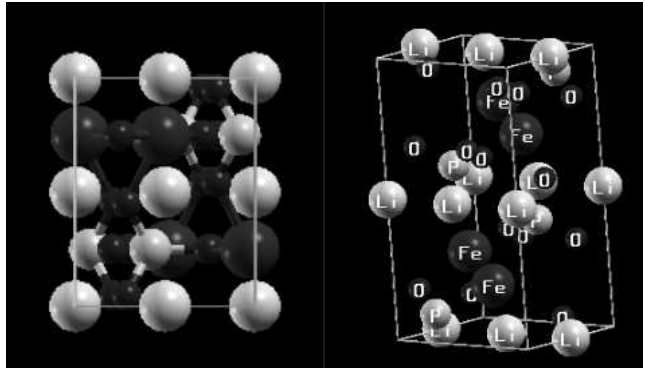} 
\caption{Schematic representations of conventional unit cell\ 
         for crystalline structure LiFePO$\rm_{4}$ in the 
         olivine structure $(a)$ 2D, and $(b) 3D$ view.
\label{fig2}}
\end{figure}

Hubbard U correction~\cite{AZA91}\ is\ applied\ 
to\ the dopant atoms.~We\ have selected\ U=4.5~eV\ 
to be optimum to our system.~Spin polarized 
calculation is allowed.~Density\ of\ states~(DOS)\ 
is\ calculated as a\ population\ of states\ 
in the spin-up and spin-down states\ at\ 
the\ chosen\ energy\ windows.~Projected\ 
DOS~(PDOS)\ is\ calculated as a\ component\ 
of\ DOS\ resolved\ onto atomic\ orbitals.\
To\ characterize\ optical\ properties,\ 
a\ dielectric\ function\ is\ computed,\ which\ 
has\ generally\ a\ complex\ nature\ $\&$\ 
is\ given\ as
\begin{equation}{\label{eq3}}
\varepsilon(\omega)={\varepsilon\rm_{1}}(\omega)\ +\ i\ 
{\varepsilon\rm_{2}}(\omega)
\end{equation}
The imaginary part\ ${\varepsilon\rm_{2}}(\omega)$\ 
is\ calculated\ from the density matrix of the\ 
electronic\ structure~\cite{HL87}\ as\ described\ 
elsewhere~\cite{GHKFB2006},\ $\&$\ given\ 
by\
\begin{equation}{\label{eq4}}
{\varepsilon\rm_{2}}(\omega)=\frac{8{{\pi}^{2}}e^{2}{\hbar}^{2}}
{\Omega {{\omega}^{2}}{m_{e}}^{2}} 
{\sum\limits_{k,v,c}}{w\rm_{k}}{{\mid}\langle{\psi\rm_{k}^{c}}
{\mid}{\bf u}{\cdot}
{\bf r}{\mid}{\psi\rm_{k}^{v}}\rangle{\mid}}^{2}
\delta(E\rm_{k}^{c}-E\rm_{k}^{v}-\hbar \omega), 
\end{equation}
where $e$ is the electronic charge,\ and $\psi\rm_{k}^{c}$\ 
and\ $\psi\rm_{k}^{v}$\ are the conduction band\ 
(CB)\ and\ valence band\ (VB)\ wave functions at k,\ 
respectively,\ $\hbar \omega$\ is the\ energy of the\ 
incident phonon,\ ${\bf u}{\cdot}{\bf r}$ is\ the\ 
momentum operator,\ $w\rm_{k}$\ is a joint\ density\ 
of states,\ $\&$\ $\Omega$\ is\ volume\ of\ the\ 
primitive cell.\ The real\ part\ ${\varepsilon\rm_{1}}(\omega)$\ 
can\ be extracted from the\ imaginary part\ 
${\varepsilon\rm_{2}}(\omega)$~(Eq.~\eqref{eq4})\ 
according to Kramer-Kronig\ relationship~\cite{FW72},\ 
as follows.\
\begin{equation}{\label{eq5}}
{\varepsilon\rm_{1}}(\omega)=1\ +\ {\frac{2}{\pi}}P
{\int\limits_{0}^{\infty}}
\frac{{\omega}'{\varepsilon\rm_{2}}({\omega}')}
{{{\omega}'}^{2}-{{\omega}}^{2}}d{\omega}'
\end{equation}
where\ $P$\ is\ a\ principal\ value.\ 
The electron energy loss function~($L({\omega})$),\
as\ given\ elsewhere~\cite{SSM2000},~is\ 
calculated~by\
\begin{equation}{\label{eq6}}
L(\omega)=\frac{{\varepsilon\rm_{2}(\omega)}}
{{\varepsilon\rm_{1}^{2}(\omega)} 
+ {\varepsilon\rm_{2}^{2}(\omega)}}
\end{equation}
The index of refraction is given by 
\begin{equation}{\label{eq7}}
n(\omega)=\frac{1}{\sqrt{2}}
{{\Bigg[}\sqrt{{{\varepsilon}_{1}^{2}}+{{\varepsilon}_{2}^{2}}} + 
{{\varepsilon}_{1}}{\Bigg]}}^{1/2}
\end{equation}
The absorption coefficient is calculated 
from dielectric function (Eq.~\eqref{eq3})\
according to 
\begin{equation}{\label{eq8}}
{\alpha}(\omega)={\sqrt{2}}\frac{\omega}{c}
{{\Bigg[}\sqrt{{{\varepsilon}_{1}^{2}}+{{\varepsilon}_{2}^{2}}} - 
{{\varepsilon}_{1}}{\Bigg]}}^{1/2}
\end{equation}
An Olivine\ structured\ within $P\rm_{nma}$,\ 
as shown in Fig.~\ref{fig2},\ is\ considered\ 
in this\ study where the\ unit cells\ 
contain 28 atoms.\ On setting up\ of this structure,\ 
literature resources~\cite{norris2014multi,li2010structural,yousefi2023first}\ 
have been\ closely followed.\
After getting the\ kinetic energy\ cut-off and the\ 
number\ of special\ k-points which give\ the best\ 
convergence\ possible of\ total energy,\ we\ 
calculated\ the total energy\ for various\ 
values of\ the lattice constants.\ Energies were\ 
calculated for\ various values of\ lattice constant,\ 
and a curve fitting\ to the values of\ total energy\ 
as a function\ of the unit cell\ volume is done\ 
according to Murnaghan equation~\cite{FDM44}.\
From the output of the curve fit,\ the values 
of bulk modulus,\ and lattice constant are\ 
predicted.\

\section{Results and Discussion \label{sec:res}}
\subsection{Structural and electronic properties}

From our DFT calculations,\ we found the values of\ 
the lattice parameters\ for $\rm{LiFePO_{4}}$ orthorhombic\ 
structure to be $a=4.67~{\AA}$. This result is\ in good\ 
agreement\ with experimental results\ in the literature,\ 
summarized in Table~\ref{tab1} below.\\
%
\begin{table}[h!]
\addtolength{\tabcolsep}{4.8mm}
\renewcommand{\arraystretch}{2.2}
\centering
\caption{The equilibrium lattice parameters [{\AA}] of 
         LiFePO$\rm_{4}$ computed from DFT (PBE) 
         and with the DFT + U (PBE + U correction),\ 
         and compared\ with available experimental 
         value.\label{tab1}}
 \begin{tabular}{lcccc}
 \hline
 \multirow{2}*{Lattice constant} & \multicolumn{2}{c}{This work} &
 \multirow{2}*{Experiment~\cite{strobridge2014characterising}} & \multirow{2}*{Error~$(\%)$}\\
 \cline{2-3}
 {} & {DFT} & {DFT + U} & {} & {}\\ 
 \hline
 $a~[{\AA}]$ & 4.76 & 4.67 & 4.71 & 0.21 \\
 $b~[{\AA}]$ & 6.00 & 5.99 & 5.94 & 0.84  \\
 $c~[{\AA}]$ & 10.28 & 10.36 & 10.35 & 0.10 \\
 \hline
\end{tabular}

\end{table}

As apparent\ from Table~\ref{tab1},\ compared to\ 
available\ experimental\ data~\cite{strobridge2014characterising},\ 
GGA\ underestimates the\ equilibrium lattice\ constant\ 
values,\ while DFT + U produces\ an optimized unit\ 
cell parameters\ which is in a\ better\ agreement\ 
with the experiment\ data.\\

\begin{figure}
\centering
\includegraphics[scale=0.5]{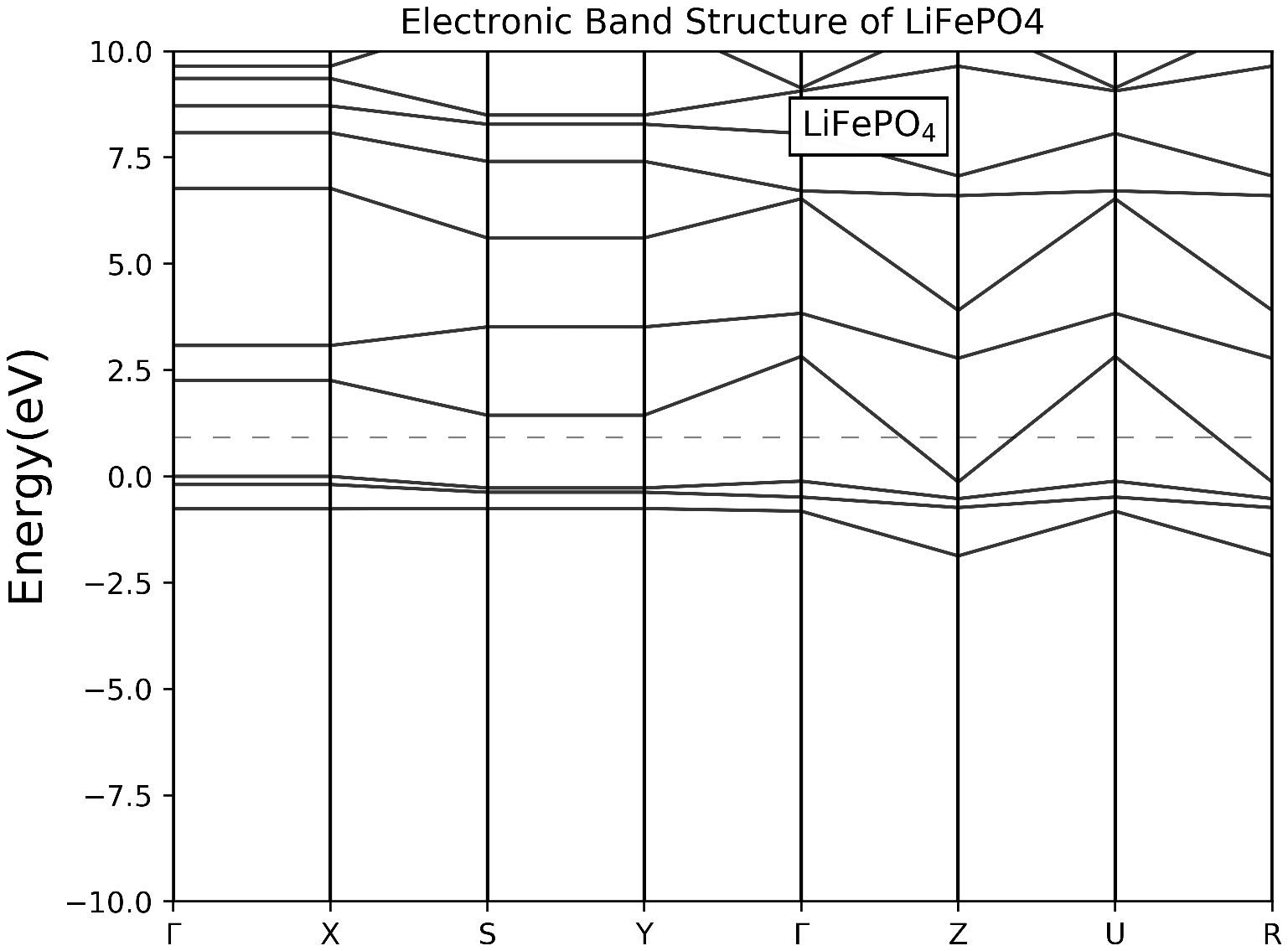}\\
\includegraphics[scale=0.4,angle=270]{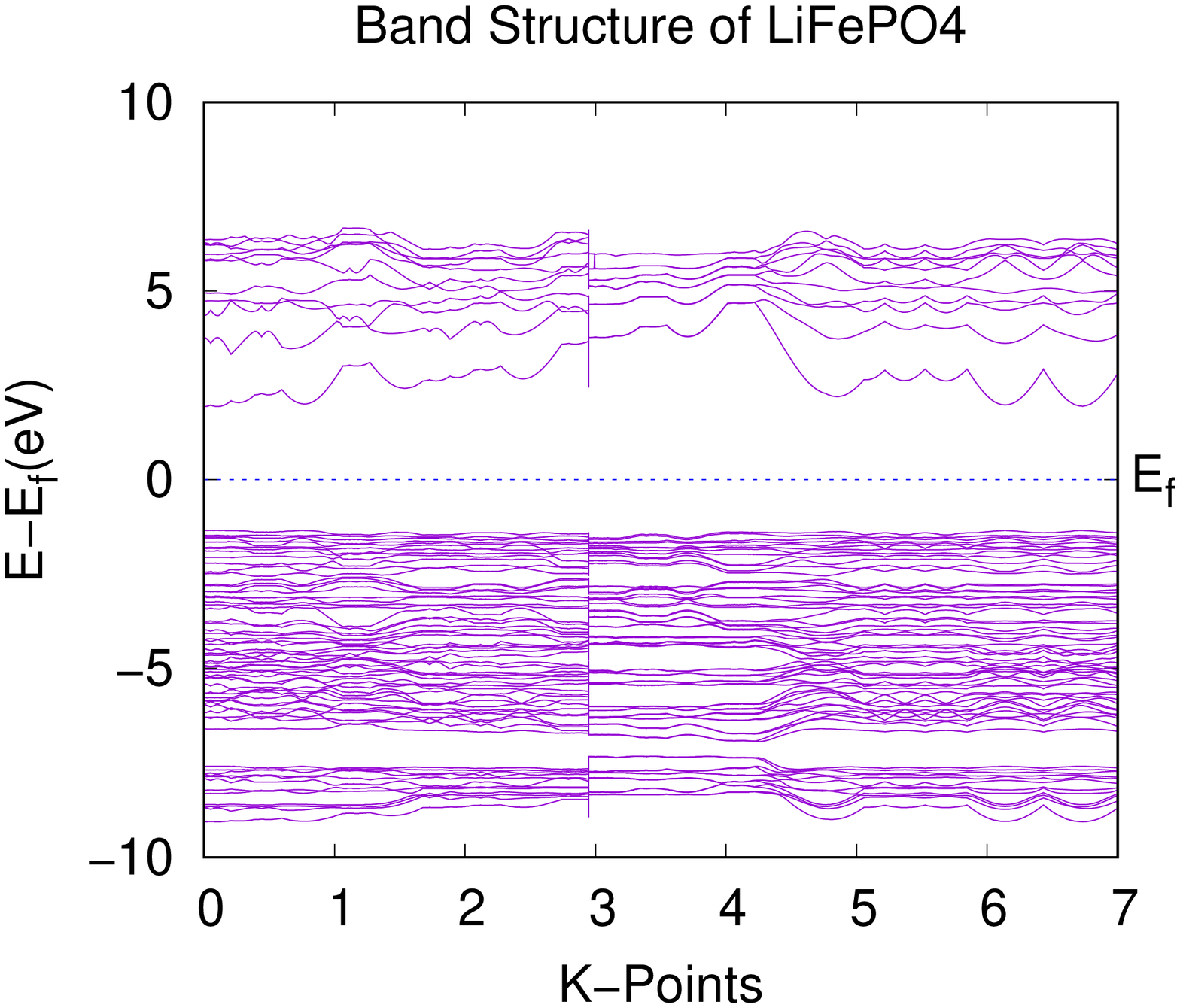}
\caption{Representations of electronic band structure of LiFePO$\rm_{4}$ 
         using (a) DFT and (b) DFT + U functionals.\label{fig3}}
\end{figure}

\begin{figure}
\centering
\includegraphics[scale=0.5]{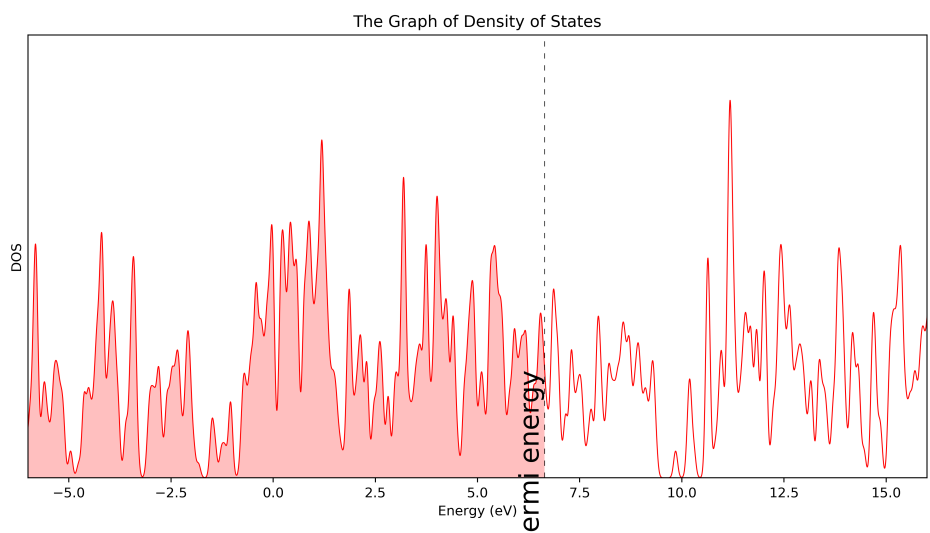}
~
\includegraphics[scale=0.5]{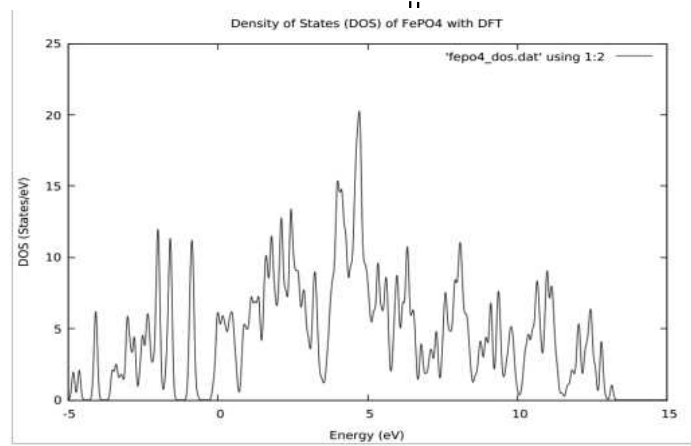} 
\caption{Density of states: (a) FePO$\rm_{4}$ and 
         (b) LiFePO$\rm_{4}$.\label{fig4}}
\end{figure}

\begin{figure}
\centering
\includegraphics[scale=0.5]{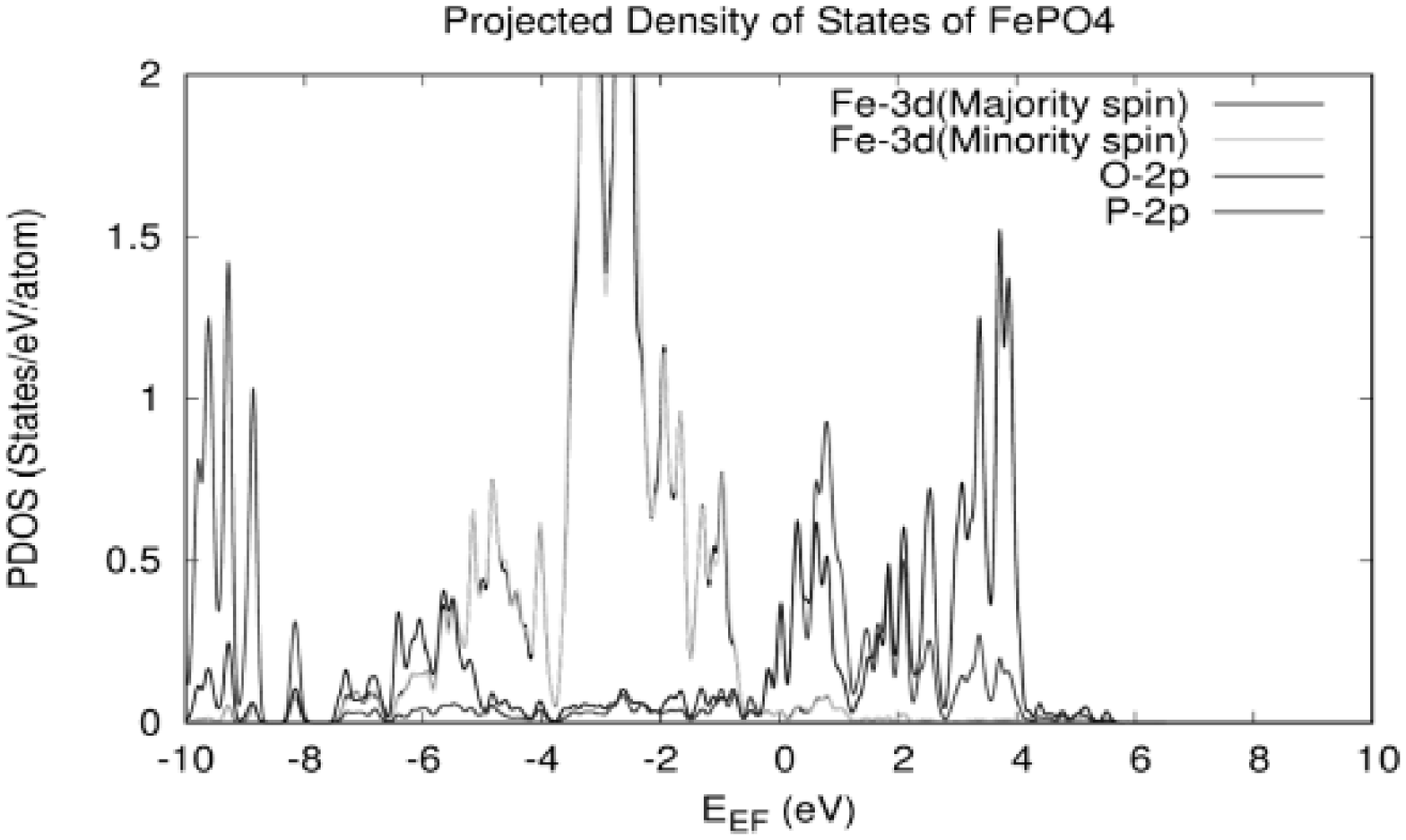}
\includegraphics[scale=0.5]{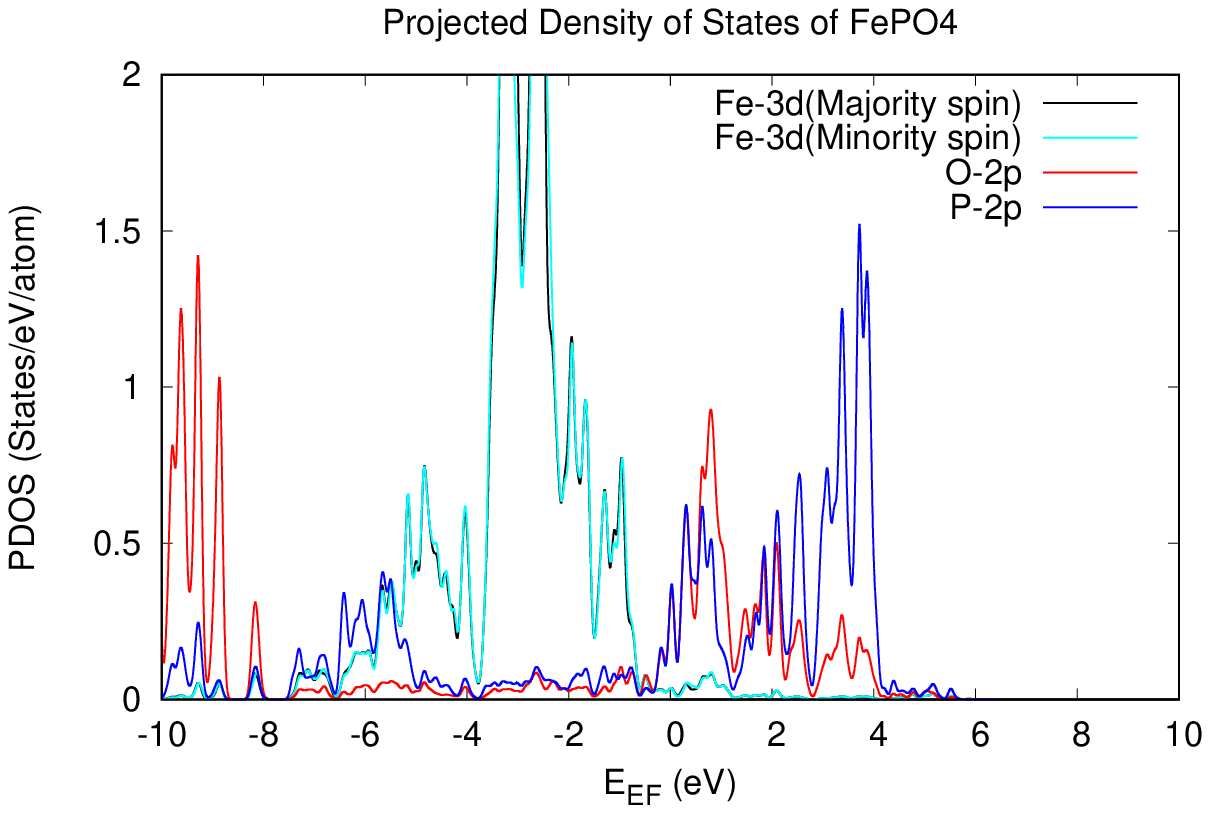}
\includegraphics[scale=0.5]{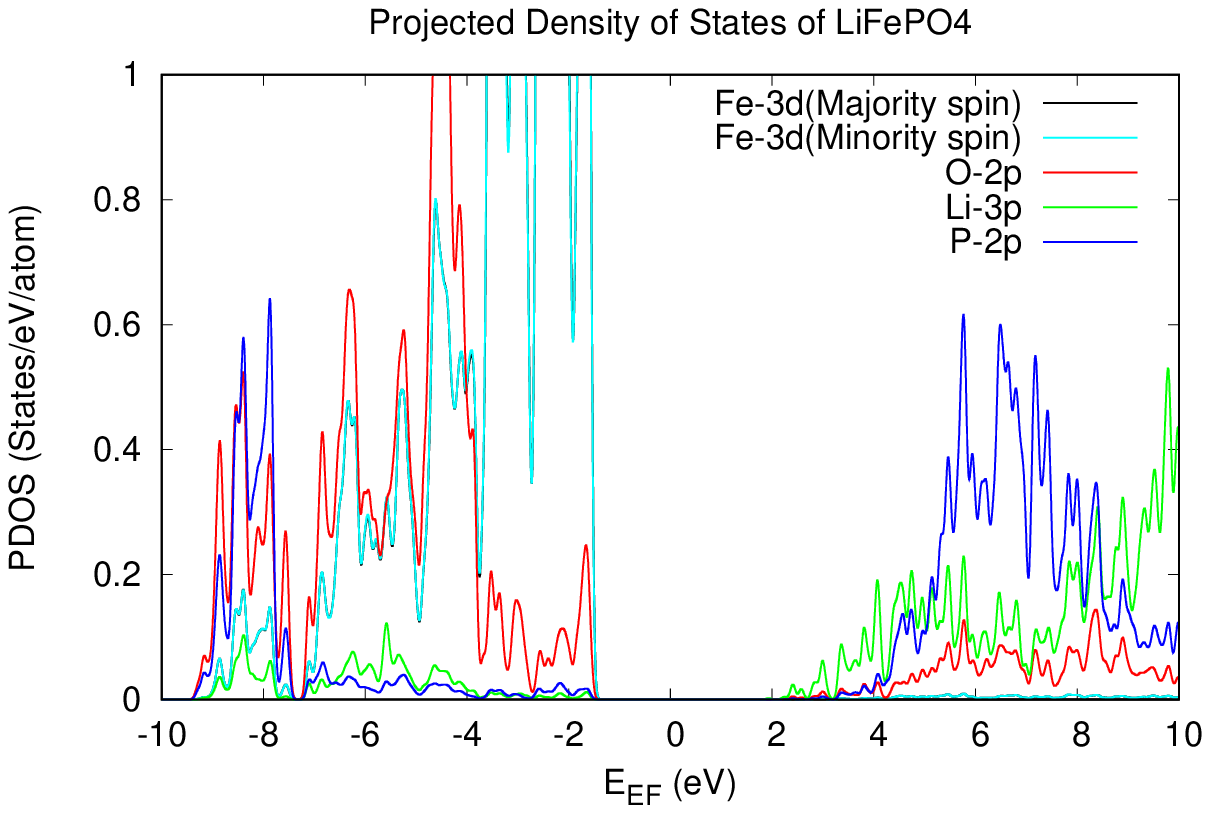} 
\caption{Projected Density of states: (a) FePO$\rm_{4}$ and 
         (b) LiFePO$\rm_{4}$ using DFT  (c) LiFePO$\rm_{4}$ 
         using DFT using DFT + U.\label{fig5}}
\end{figure}

The band structure\ for ferromagnetic forms of\ 
FePO$\rm_{4}$ and LiFePO$\rm_{4}$ are shown\ 
in\ Fig.~\ref{fig3}.\
A direct band gap seems 0.10~eV with DFT (GGA), and\ 
3.82~eV using DFT + U formalism for LiFePO$\rm_{4}$.\ 
Experimental band gap\ results for LiFePO$\rm_{4}$\ 
are 2.86-4.00 eV,\ as reported in a literature~\cite{yang2012first}.\ 
Thus,\ while the GGA (PBE) functional predicts a\ 
metallic behavior with the Fermi level of the system\ 
crossing the\ minority-spin $d$ states,\ DFT + U\ 
is effective in\ predicting a\ correct band gap.\ 
In the total and projected DOS analysis illustrated\ 
in Figs.~\ref{fig4}~$\&$~\ref{fig5},\ the main\ 
contribution to\ valence band\ was associated with\ 
O 2$p$ states\ with minor contributions\ from Fe-3$d$\ 
states.\ At the conduction band,\ the\ dominant\ 
contribution\ is by Fe-3$d$ atomic orbitals\ 
with a small content\ oxygen atomic orbitals.\ Thus,\ 
it is possible\ to assume that\ an electron transfer\ 
inside the\ band gap region\ should occur between\ 
2$p$ orbitals\ of oxygen atom\ and 3$d$ orbitals\ 
of the Fe in\ tetrahedral configurations,\ 
represented by\ FePO$\rm_{6}$ clusters.\

\begin{figure}
\centering
\includegraphics[scale=0.43]{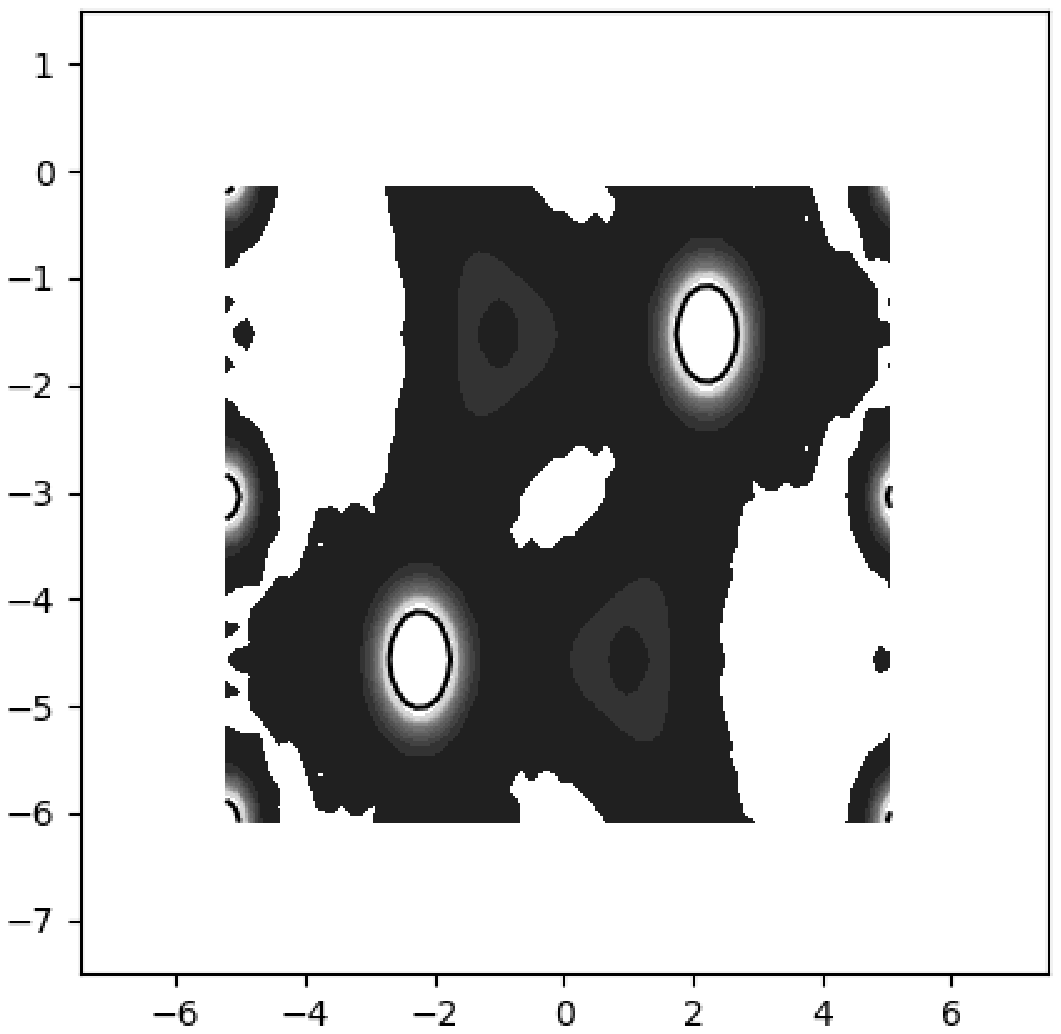} 
\includegraphics[scale=0.2]{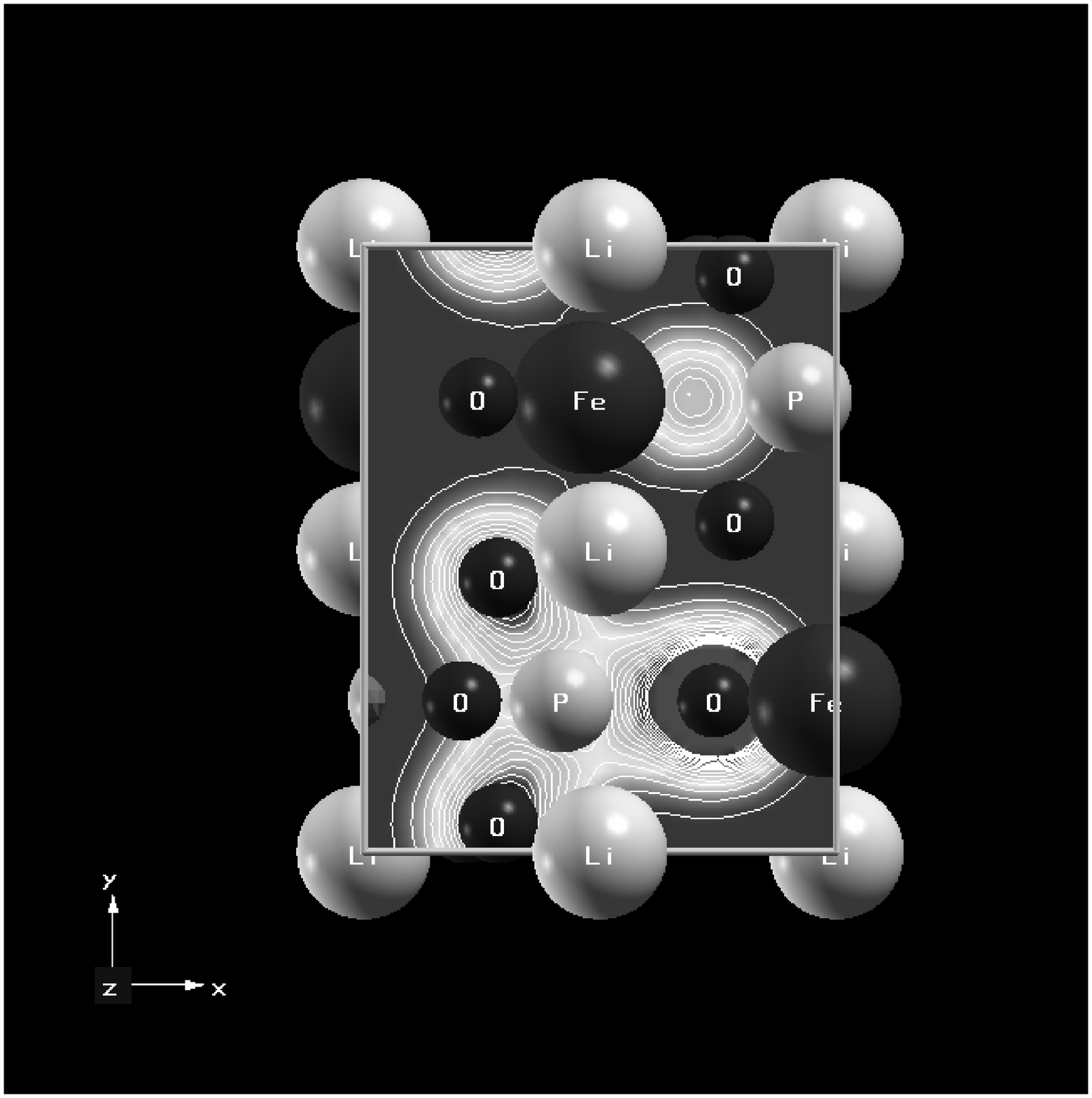}
\includegraphics[scale=0.2]{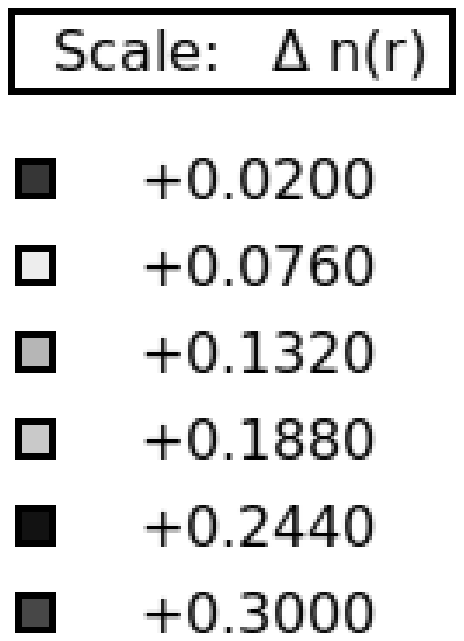}  
\caption{Charge density plot (a) The total charge density plot, 
         and (b) The electronic charge density contour plots 
         for majority electrons in LiFePO$\rm_{4}$ in 
         [001] direction, (c) The Thermo scale.\label{fig6}}
\end{figure}

In order to know\ the distribution\ of the total\ 
charge density\ of LiFePO$\rm_{4}$ orthorhombic\ 
structure,\ we have calculated\ the charge density\ 
distribution.\ From the result\ we can observe\ 
that LiFePO$\rm_{4}$\ structure makes\ a covalent\ 
bonding.\ From Fig.~\ref{fig6},\ it is clear that\ 
in LiFePO$\rm_{4}$ structure\ Fe-Fe shows a very\ 
weak charge density\ but when we move\ to P-P\ 
bonding,\ there is stronger\ charge density.\ 
Also as clear as\ it is from the scale,\  
purple color\ shows the greater\ charge density\ 
than the remaining atoms.\\
Using Bader decomposition~\cite{RB90},\ which uses stationary\ 
points\ in the bulk\ electron density\ to partition electrons\ 
among different\ atoms,\ and within\ the approaches adopted\ 
in a literature~\cite{TSH2009},\ we have calculated\ the Bader\ 
charges on\ each atom\ which are\ contained within a unit cell\ 
of LiFePO$\rm_{4}$,\ as given in Table~\ref{tab2}.\
\begin{table}[h!]
\addtolength{\tabcolsep}{2.8mm}
\renewcommand{\arraystretch}{2.2}
\centering
\caption{Bader Charges values on each atom in LiFePO$\rm_{4}$,\ 
         where Z means atomic number.\label{tab2}}
\begin{tabular}{lcccccc}
\hline
{} & {} & \multicolumn{3}{c}{Results of Bader Charge analysis} & {} & {}\\
\cline{3-5}
 Atom & {Z} & This work & Calc. & Ref.~\cite{kuriplach2019first} & Error $(\%)$ & Nominal charge\\
 \hline
 Li & 3.0 & 2.12 & +0.90 & +1.00 & 10 & +1 \\
 Fe & 16.0 & 14.5 & +1.50 & +1.55 & 3.2 & +2 \\
 P & 5.0 & 0.12 & +4.88 & +5.00 & 2.4 & +5 \\
 O & 6.0 & 7.83 & -1.83 & -1.83 & 0.0 & -2 \\
 O & 6.0 & 7.87 & -1.87 & -1.89 & 1.1 & -2 \\
 O & 6.0 & 7.90 & -1.90 & -1.92 & 1.0 & -2 \\
 \hline
\end{tabular}
\end{table}
The deviation\ from the ideal ionic\ charge density\ 
is more significant\ for Li than for Fe,\ suggesting\ 
a higher degree\ of covalent in $\rm{Li-O}$\ than in the\ 
$\rm{Fe-O}$ interaction.\ The homogeneous distribution\ 
of contour lines\ represents the\ strongly covalent\ 
character in\ the interaction of\ the Li and Fe\ 
cations atoms\ with oxygen anion\ on the [001]\ 
analyzed plane.\ The observed behavior\ occurs\ 
because of the\ hybridization between\ the O 2$p$\
atomic orbitals\ with Fe 3$d$\ atomic orbitals.\ 
For FePO$\rm_{4}$,\ the Fe states\ are well\ 
hybridized with\ O 2$p$ states\ throughout\ 
the valence band.\ This is shown both in\ 
the projected\ densities of states plot\ 
of Fig.~\ref{fig5} and in the contour plot\ of 
Fig.~\ref{fig6}.
\subsection{Optical properties}
The optical property\ of matter can be described\ 
by the knowledge\ of the complex\ dielectric function,\ 
which describes\ the optical response\ of the\ 
material to\ the external\ electromagnetic\ 
field~\cite{miara2015first}.
The imaginary part ($\varepsilon_2(\omega)$) of the\ 
dielectric function\ implies the optical transition\ 
mechanism.\ Each peak\ in the imaginary\ part of the\ 
dielectric function\ corresponds to\ an electronic\ 
transition.\ The imaginary part\ of the complex\ 
dielectric function\ is related\ to a measure\ 
of optical absorption.\ The real part\ of dielectric\ 
function\ is obtained\ using Eq.~\eqref{eq5}\
and describes\ other properties\ such as optical\ 
transmission.\ 

\begin{figure}
\centering
\includegraphics[scale=0.5]{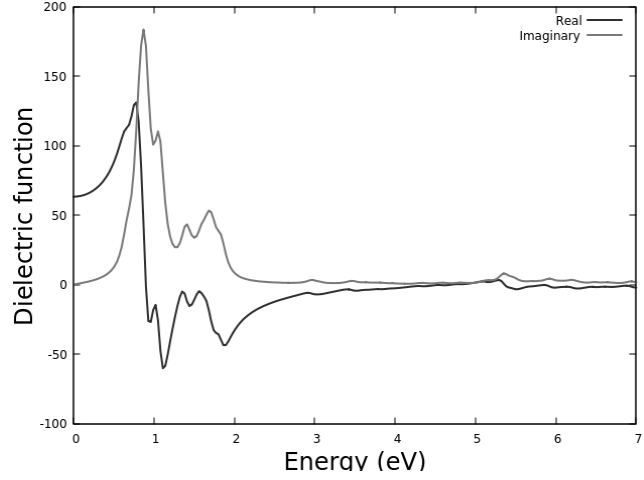} 
\caption{Dielectric function of LFP.\label{fig7}}
\end{figure}

Figure~\ref{fig7} depicts\ a graph of dielectric\ 
function against\ the photon energy\ which gives\ 
the calculated\ results of the\ real $(\varepsilon_{1})$\ 
and imaginary\ $(\varepsilon_{2}$) parts of the\ 
dielectric functions\ which are\ connected by\ 
the dispersion\ relations~\cite{ceder2011recharging}.\ 
The static value\ of $\varepsilon_{1}(0)$\ is 64.07,\ 
and it reaches\ a maximum value\ of 130.27\ at photon\ 
energy of 0.8 eV.\ With increasing\ photon energy,\ 
it gradually\ decreases to\ a minimum\ value of -58.98\ 
at photon energy of\ around 1.12~eV,\ before it starts\
to slightly\ increase again.\ 
The distinctive\ features (peaks)\ of $\varepsilon_{2}$\ 
are due to\ optical transitions\ involving hybrid O-2$p$\ 
and Fe 3$d$ orbital,\ as is the case in $\rm{LiFePO_4}$ (LFP)~\cite{ng2019first}.
Interestingly in LFP,\ the band transitions\
seem to happen\ without excitonic effects.\

\begin{figure}
\centering
\includegraphics[scale=0.25]{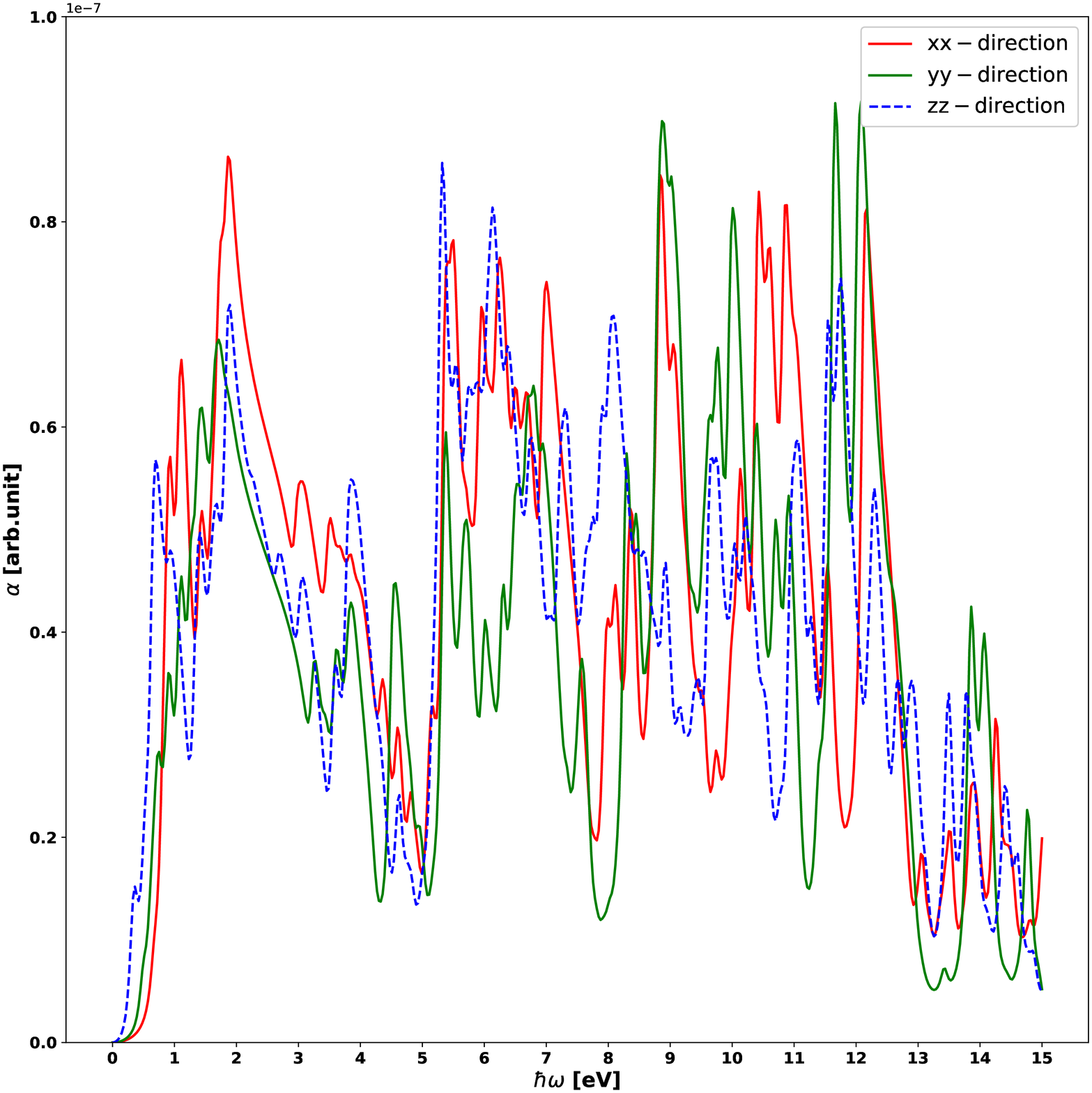} 
\includegraphics[scale=0.25]{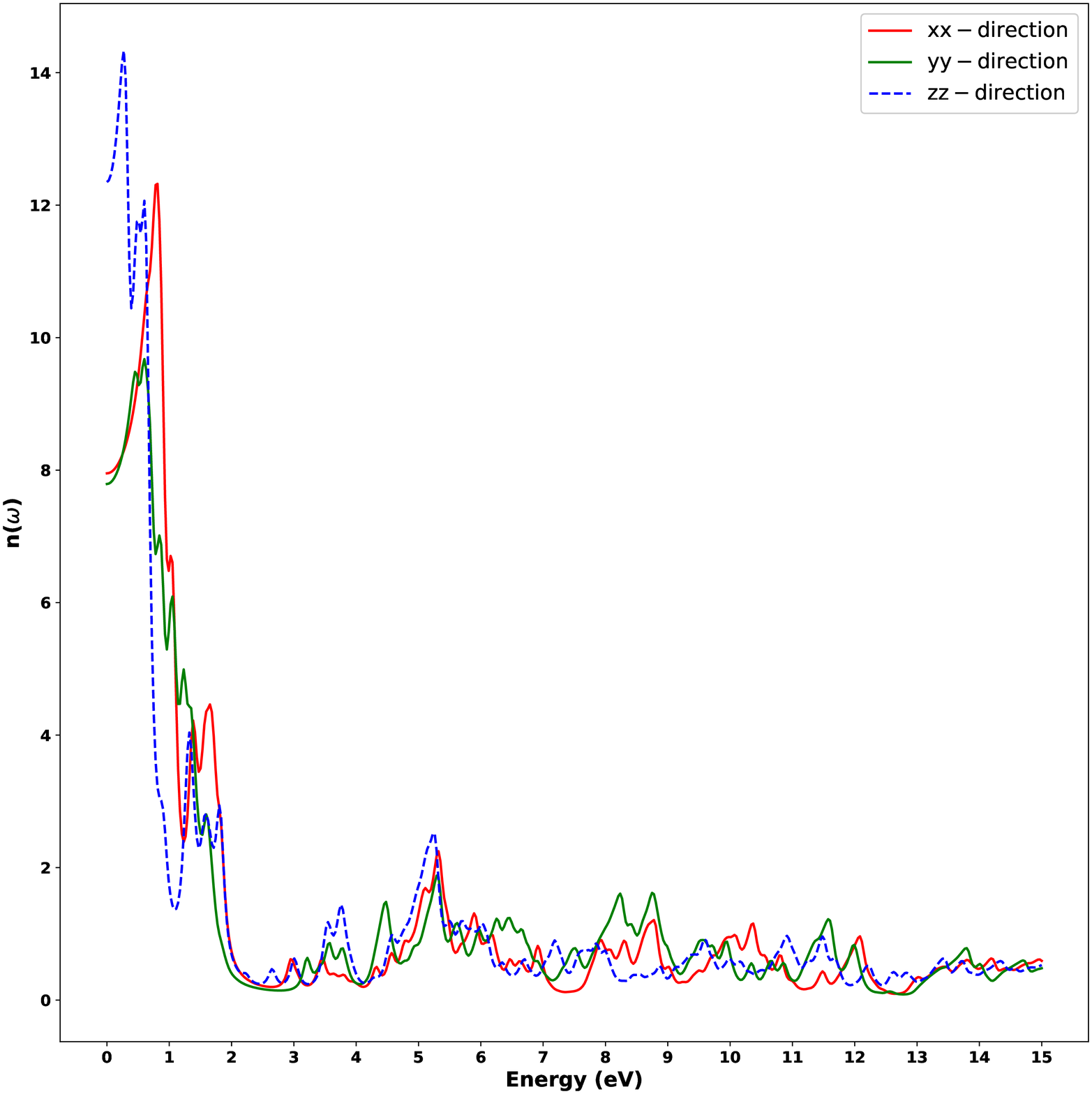} 
\includegraphics[scale=0.25]{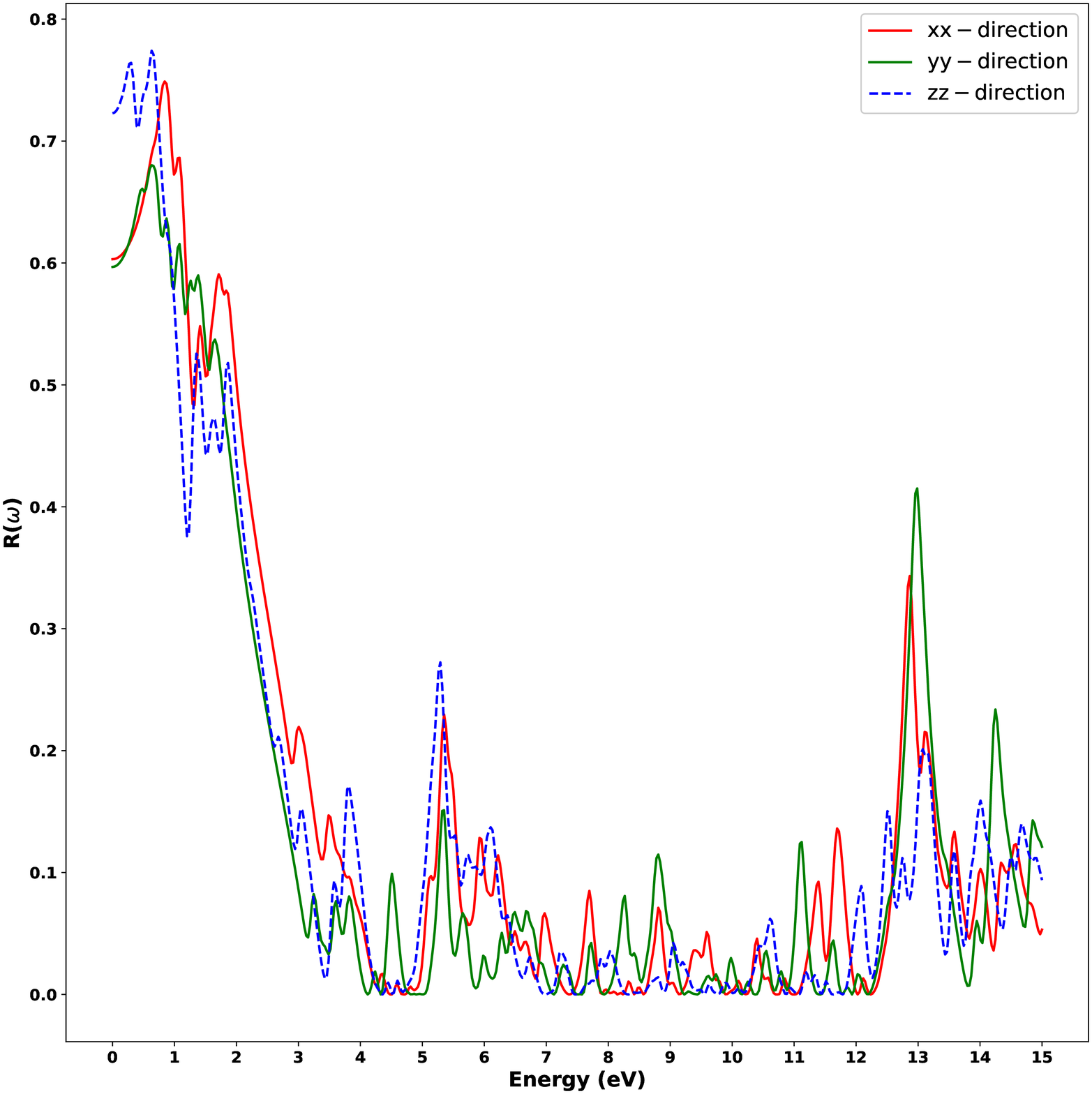}
\caption{The graph of: (a) absorption coefficient,\ 
         (b) refractive index, and (c) reflectivity,\ 
         of LFP.\label{fig8}}
\end{figure}

The absorption\ coefficient\ determines which\ 
light of a particular\ wavelength\ is absorbed\ 
by a material~\cite{ong2011first}.\ 
In a material\ with a low absorption\ coefficient,\ 
light is\ only poorly absorbed,\ and if the material\ 
is thin enough,\ it will appear\ transparent to that\ 
wavelength.\ The absorption\ coefficient\ depends on\ 
the material and\ also on the wavelength\ of light\ 
which is\ being absorbed.\ 
The absorption\ coefficient of\ LiFePO$\rm_{4}$ is\ 
given\ in Fig.~\ref{fig8}.\ At 2.5~eV,\ an 
absorption peak in the $xx$-direction\ is noticed.\ 
At 5.8~eV,\ an absorption peak\ in the $zz$-direction\ 
is noticed.\ At 12.5~eV,\ an absorption\ peak in the\ 
$yy$-direction\ is\ noticed.\ 

The refractive index\ computed using Eq.~\eqref{eq7}\
is shown\ in Fig.~\ref{fig8}.\ 
The maximum refractive\ index value of 14.1 occurs\ 
at photon energy of 0.5~eV.\ The index of refraction\ 
at zero\ photon energy\ is $n(0)=7.9$.\ 
Between photon\ energies of\ 0 and 2.0 eV,\ the\ 
index of\ refraction attains\ maximum and then\ 
gradually\ decreases to\ $n({\omega})=1$.\  
Reflectance is ability of\ a substance to reflect\ 
radiation.\ As shown in Fig.~\ref{fig8},\
the reflectivity at\ zero photon energy\ has\ 
values of 0.61\ in the $xx$-direction,\ 0.59\ 
in the $yy$-direction,\ and 0.72 in the $zz$-direction.\ 
At photon\ energy of 1.84 eV,\ the highest\ 
reflectivity peak\ of 0.75 is\ noticed in the\ 
$xx$-direction.\ At photon\ energy of 1.70 eV,\ 
a highest reflectivity\ peaks of 0.78 in\ the\ 
$zz$-direction and 0.68 in the $yy$-direction\
is\ noticed.\\

\begin{figure}
\centering
\includegraphics[scale=0.5]{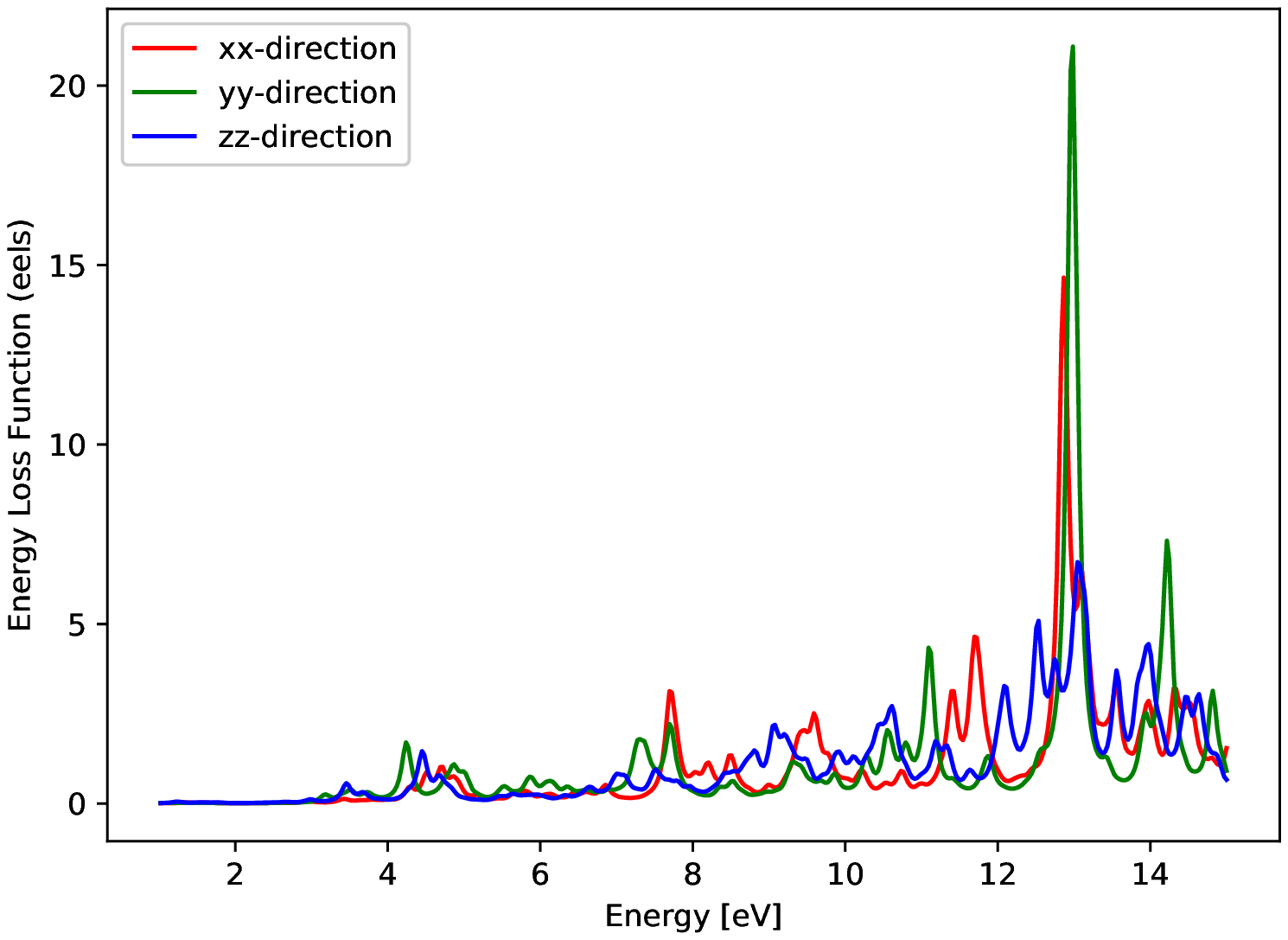} 
\caption{The electron-energy loss spectrum of LFP.\label{fig9}}
\end{figure}

The electron energy\ loss spectrum\ (eels) displays\ 
a prospect of\ a material resulting\ in some of the\ 
electrons\ undergoing\ inelastic scattering,\ 
which means\ that they\ lose energy\ and have\ 
their paths\ slightly and\ randomly deflected.\
The eels for the LFP system is shown in Fig.~\ref{fig9}.\
At 13.5 eV,\ we have\ the highest energy loss in\ 
the $xx$-direction.\ 

\begin{figure}
\centering
\includegraphics[scale=0.5]{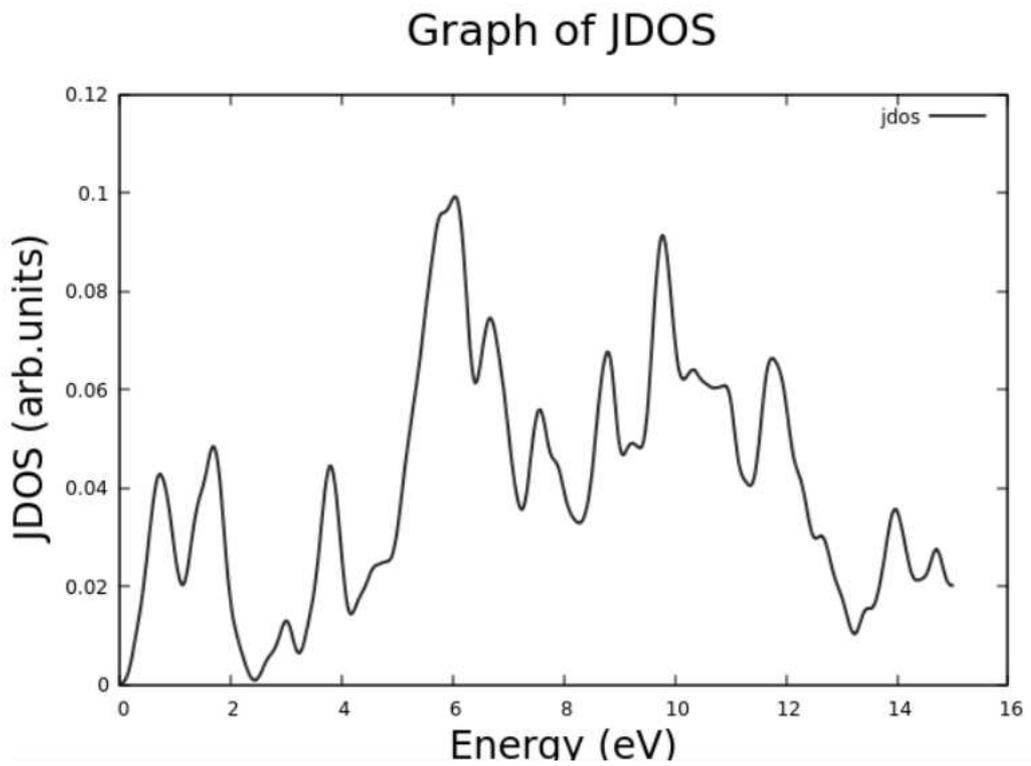} 
\caption{The joint density of states (JDOS) of LFP.\label{fig10}}
\end{figure}

The Joint\ density of\ states (JDOS)\ is an indicator\ 
of the number of\ available states for\ photons to\ 
interact with.\ For optical\ absorption process,\ 
it is an\ important part\ of optical characteristics\ 
of a given material.\ The JDOS of\ LFP shows the\ 
sharpest peak at 6.0~eV (Fig.~\ref{fig10}).
\section{Conclusion\label{sec:conc}}

In this study,\ we have\ employed a DFT + U\ 
calculations to\ investigate the\ structural,\ 
electronic,\ optical,\ and magnetic properties\ 
of LiFePO$\rm_{4}$\ cathode material\ for Li-ion\ 
batteries.\ We have performed\ the structural\ 
optimization and\ calculated the\ equilibrium\ 
parameters such\ as the lattice\ constants,\ 
and the bulk modulus\ using QE code and found\ 
that $a=4.76~{\AA}$, $b=6.00~{\AA}$, $c=10.28~{\AA}$,\ 
and $\beta = \rm{90.2~GPa}$.\ The results obtained\ 
are in agreement\ with experimental\ results reported\ 
in the literature.\\
	
The result obtained\ with a\ DFT + U\ 
showed that\ LiFePO$\rm_{4}$\ is direct band gap\ 
materials with\ a band gap of\ 3.82~eV,\ which is\ 
within a range\ of the\ experimental values.\ We have\ 
analyzed the\ projected density\ of states which\ 
suggest that\ the majority spin\ states of FePO$\rm_{4}$\ 
have substantial\ covalent character\ due to the\ 
energetic overlap\ of the O states\ with the Fe states.\ 
In LiFePO$\rm_{4}$,\ there is less\ covalent character\ 
such that\ the Fe states form\ narrow bands above\ 
the O bands with\ a relatively lower\ extent of\ 
mixing.\ Thus,\ based\ on the results,\ it seems\ 
that\ LiFePO$\rm_{4}$\ is more stable\ than FePO$\rm_{4}$.\\

On the basis of the\ predicted optical absorbance,\ 
reflection,\ refractive index,\ and energy loss function,\ 
LiFePO$\rm_{4}$ seems to be viable and cost-effective\ 
as a cathode material\ for\ Li-ion battery.\ 
Furthermore, it appears that\ the DFT + U\ 
formalism is\ the most suitable\ choice to investigate\ 
the strongly\ correlated\ LiFePO$\rm_{4}$ system,\ 
contributing\ to further\ literature resource\ 
involving\ such\ technological material.\

\section*{CRediT authorship contribution statement}
A.K.~Wabeto\ conducted the\ DFT calculations,\
and wrote the draft manuscript;\ K.N.~Nigussa\ 
directed the\ research process\ and carried out\ 
the writing of the revised\ manuscript;\ and\  
L.D.~Deja supported\ on the research process.\
\section*{Declaration of Competing Interest}
The authors declare\ that they have no known\ 
competing financial\ interests or personal\ 
relationships that\ could have appeared\ 
to influence the\ work reported in this paper.\

\section*{Acknowledgements}
We are grateful to the Ministry of Education\ 
of\ Ethiopia for financial support.\ The\ 
authors also acknowledge\ the\ Department of\ 
Physics at\ Addis Ababa University.\ 
The~office~of~VPRTT~of Addis\ Ababa\ University\ 
is also warmly~appreciated~for\ supporting~this\ 
research under a\ grant~number~AR/053/2021.\ 

\section*{Data\ Availability\ Statement}
The data that\ support the findings\ of\ 
this study\ are available\ upon reasonable\ 
request\ from the\ authors.\

\section*{ORCID\ iDs}
K.N.\ Nigussa.\\
\url{https://orcid.org/0000-0002-0065-4325}.
\section*{References}
\bibliographystyle{elsarticle-num}
\bibliography{refs.bib}
\end{document}